# Comment on "Field-dependent surface resistance for superconducting niobium accelerating cavities" published in

Wolfgang Weingarten[1]

The purpose of this comment is two-fold. First, I am responding to a criticism I have received about the paper of ref. 1. Second, I also intend to supplement statements in ref. 1 in the light of new results obtained since then and presented in papers refs. 2, 3 and also in papers referenced below.

The aim is to show the important role of small (compared to coherence length) weak superconducting (sc) spots when located at the surface in connection with the proximity and percolation effects.

The criticism concerns the "Cooper limit" formula for the superconducting coupling constant $(NV)_{eff}$. This formula describes the boundary effect for the mixture of niobium and niobium monoxide in close proximity,

$$(NV)_{eff} = \frac{(NV)_N N_N v_N + (NV)_S N_S v_S}{N_N v_N + N_S v_S}. \quad (1)$$

Here $(NV)_S$ is the sc coupling constant of the "strong" (S) superconductor Nb, $(NV)_N$ is that of the "weak" (N) superconductor NbO, and $v_S$ and $v_N$ are the volume densities of the respective components S and N within the mixture. $N_S$ and $N_N$ are the "electron density of states" (DOS), in units of eV$^{-1}$m$^{-3}$. In ref. 1, $N_S$ and $N_N$ were intentionally referred to as the "electron densities" of the respective superconductors in units of m$^{-3}$, so that they were claimed to be numerically identical to the DOS, but no justification was provided. This deficiency is now to be corrected. However, this change has no significant effect on the value of $(NV)_{eff}$, as will be shown below.

In the following, $(NV)_{eff}$ is re-evaluated using data that appeared after the publication of ref. 1 [4, 5]. Note that neither the absolute values of $N_S$ and $N_N$ nor those of the volume densities $v_S$ and $v_N$ are critical, but their relative values, cf. eq. 1. From $(NV)_{eff}$ follows the critical temperature $T_{cNS}$ of the mixture Nb/NbO as a function of the relative composition $x$:

$$T_{cNS} = 1.14\, \Theta_D e^{-1/(NV)_{eff}} \quad ,$$

shown in Fig. 1. $\Theta_D$ is the Debye temperature of the more abundant component of the mixture, NbO ($\Theta_D$ = 472 K). The data used are summarized in Table 1 and 2.

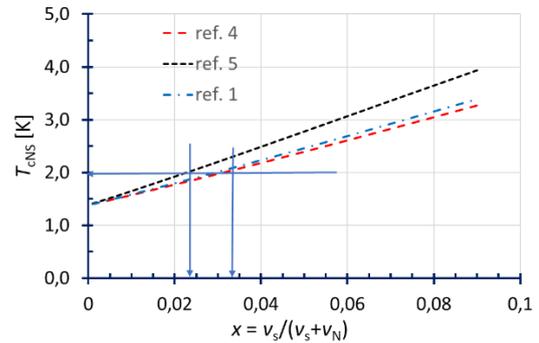

Fig. 1: The critical temperature $T_{cNS}$ of the Nb/NbO mixture vs the relative composition $x = v_S/(v_S+v_N)$ of the S (Nb) component

Table 1: Electron density $n$ of Nb and NbO

|  | Nb | NbO | unit |
|---|---|---|---|
| $A$ | 92,9 | 108,9 | g |
| $\rho_m$ | 8,57 | 7,3 | g/cm$^3$ |
| $n$ | 5,55·10$^{22}$ | 4,04·10$^{22}$ | atoms/cm$^3$ |
| | $n = 0{,}6022 \cdot 10^{24} \cdot \rho_m/A$ | | |

$A$: atomic weight, $\rho_m$: mass density

---

[1] wolfgangweingarten@t-online.de

It can be seen from Fig. 1 that the experimentally determined percolation temperature $T^* = 2$ K [1] corresponds to a critical mixture $x^*$ in the interval of 2.3 % to 3.3 %. This interval includes the critical mixture of ref. 1 ($x^* = 3{,}01$ %).

Table 2. Density of states (DOS) of Nb and NbO

| DOS | Nb (S) | NbO (N) | unit | $N_N/N_S$ |
|---|---|---|---|---|
| Ref. 4 | 1,3 | 0,55 [1)] | states/(Nb·eV) | |
| | 7,22·10$^{22}$ | 2,22·10$^{22}$ | atoms/(eV·cm$^3$) | 0,31 |
| Ref. 5 | 1,8 | 0,4 | states per eV | 0,22 |
| Ref. 1 | 5,56·10$^{22}$ | 1,60·10$^{22}$ | electrons/(eV·cm$^3$) | 0,29 |

[1)]This number concerns the $Nb_3O_3$ structure, not the hypothetical $Nb_4O_4$ structure

Since the publication of ref. 1 another paper dealing with percolation through voids around overlapping spheres appeared [6], as a supplement to the one in ref. 1 [7]. This paper gives a wider interval for the critical mixing, about 2.5 % to 4.5 %, which agrees quite well with the above results. Therefore, these results do not contradict ref. 1.

Another conjecture of ref. 1 concerns the gain of diamagnetic (surface) energy at the expense of the loss of condensation energy, when a sufficiently small weak sc spot on the surface transforms into a normal conducting (nc) one under the action of the magnetic field. This effect is visualized in the commonly plotted quality factor $Q$ vs the magnetic field $B$, as shown in Fig. 2, based on a closed form (except for the low field dependence of Q on $B < 20$ mT, the details of which will be explained later).

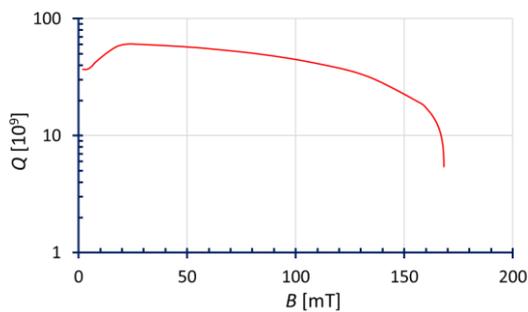

Fig. 2: Q vs magnetic surface field $B$ as obtained with numbers as of ref. 1 ($T = 2.5$ K, geometry factor $G = 250\ \Omega$)

The numbers used for this plot are those from ref. 1, "collective data". The plot shows the quite commonly observed dependence of the Q-value on the magnetic field, usually referred to as "medium field slope" and "Q-drop".

The assumption of small weak sc spots also points the way to explaining two other experimental results. One concerns the enhancement of the Q-value by doping with nitrogen [8], the other allows to explain the dependence of the Q-value on the external static magnetic field, as observed in superconducting cavities whose inner surface is covered with a thin niobium layer [9].

The first observation is described in ref. 2. The explanation is again based on a binary uniform metallic mixture of niobium and nitrogen as a weak superconductor subject to the proximity effect. This weak superconductor is embedded in high quality niobium metal. At very small field $B$, the Q-value remains constant up to the critical field $B^*$ of the weak superconductor ($\approx 5$ mT). As $B$ increases above $B^*$, different RF losses appear near the characteristic volume fraction (percolation threshold) of the nitrogen component inside the weak superconductor. They are either increasing or decreasing. The variation of the surface losses there depends on the frequency and the depth of the nitrogen-doped surface layer. The latter determines the saturation field $B_c^*$ ($\approx 20$ mT) of the weak superconductor, above which its Q-value remains constant. It is interesting to note that the characteristic increase or decrease of the Q-value at low fields is also observed in undoped cavities, indicating a rather general phenomenon. This fact proves the importance of the percolation effect as a geometry phenomenon.

This last remark suggests that the often-observed Q-increase (sometimes Q-decrease) in niobium bulk cavities could have the same origin caused by the contamination surface layer which is thinner than that of the nitrogen doped one. This conjecture could complement the explanation for the Q increase at low field (due to latent heat) given in ref. 1. The typical Q vs B plot of Fig. 2 accounts for this statement, as can be seen for $B < B^* \approx 20$ mT.

The second observation is given in ref. 3. The magnetic flux is trapped in weak sc spots that have turned into nc fluxons, as has been known for a long time. A new feature is the occurrence in nitrogen-doped cavities, whose response to static magnetic flux has a characteristic maximum [10]. This is explained by the fact that weak sc spots have a different surface resistance than the rest of the niobium surface and therefore build up a space charge when current flows through them. Thus, they act like a capacitor in an alternating field. Together with the inductance formed by the superconductor, they thus form an LC resonant circuit. A lumped circuit model is used to determine the associated resonant frequency, which, with the aid of a fit routine, leads to new results on the properties of the weak sc spots. Compared to standard niobium, they exhibit lower critical temperature and electron density, indicating dirty and/or disordered niobium with many dislocations or dissolved oxygen near the solubility limit.